\documentclass[aps,prl,reprint,groupedaddress, showpacs]{revtex4-1}

\usepackage{graphicx}
\usepackage{epstopdf}
\usepackage{amssymb}
\usepackage{amsmath}
\usepackage{bm}
\usepackage{braket}

\begin{document}

\title{Collective emission of matter-wave jets from driven Bose-Einstein condensates} 

\author{Logan W. Clark, Anita Gaj, Lei Feng, Cheng Chin}
\affiliation{James Franck Institute, Enrico Fermi Institute and Department of Physics, University of Chicago, Chicago, IL 60637, USA}

\date{\today}

\begin{abstract}
Scattering is an elementary probe for matter and its interactions in all areas of physics. Ultracold atomic gases provide a powerful platform in which control over pair-wise interactions\cite{Chin2010} empowers us to investigate scattering in quantum many-body systems\cite{Bloch2008}. Past experiments on colliding Bose-Einstein condensates have revealed many important features, including matter-wave interference\cite{Andrews1997,Anderson1998}, halos of scattered atoms\cite{Chikkatur2000,Buggle2004}, four-wave mixing\cite{Deng1999, Vogels2002}, and correlations between counter-propagating pairs\cite{Perrin2007, Perrin2008, Jaskula2010}. However, a regime with strong stimulation of spontaneous collisions\cite{Pu2000, Duan2000, Vardi2002,Bach2002,Zin2005,Norrie2005,Ogren2009,Deuar2014,Wasak2014,RuGway2011} analogous to superradiance\cite{Inouye1999, Moore1999,Schneble2003} has proven elusive. Here we access that regime, finding that runaway stimulated collisions in condensates with modulated interaction strength cause the emission of matter-wave jets which resemble fireworks. Jets appear only above a threshold modulation amplitude and their correlations are invariant even as the ejected atom number grows exponentially. Hence, we show that the structures and occupations of the jets stem from the quantum fluctuations of the condensate. Our findings demonstrate the conditions for runaway stimulated collisions and reveal the quantum nature of the matter-wave emission.
\end{abstract}

\maketitle

\begin{figure*}
	\centering
	\includegraphics[width=170mm]{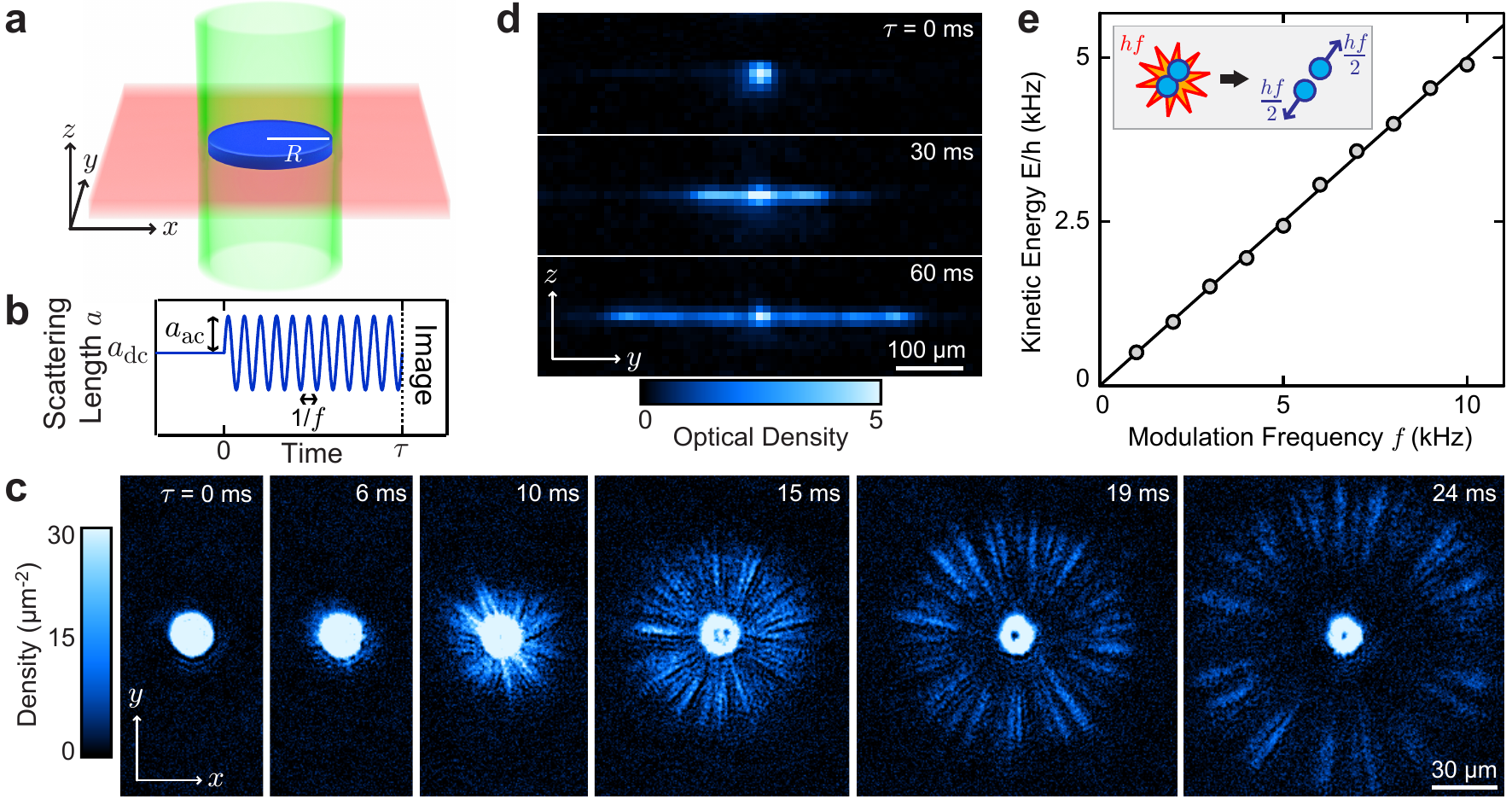} 
	\caption{\textbf{Two-dimensional emission of matter-wave jets from Bose condensates with modulated interactions.} \textbf{a.} Disk-shaped Bose-Einstein condensates (blue) of radius $R$ are trapped at the intersection of two lasers, which create a repulsive cylindrical shell (green) and an attractive sheet (red). 
	\textbf{b.}  In a typical experiment we modulate the scattering length as $a(t) = a_{\mathrm{dc}}+a_{\mathrm{ac}}\mathrm{sin}(\omega t)$ where $\omega\equiv2\pi f$ for a time $\tau$ before collecting an image of the resulting density distribution.
	\textbf{c.} Top-down images of condensates after modulating the scattering length at $f=3.5$~kHz and $a_{\mathrm{ac}}=60~a_0$ show condensates ejecting a sudden burst of narrow jets. Note that the internal structure of the remaining condensate suffers from imaging artifacts due to the extremely high density.
	\textbf{d.} Side-view images of the condensates taken at $f=3.5$~kHz show atoms emitted predominantly in the horizontal plane. \textbf{e.} The measured kinetic energy per emitted atom for a range of modulation frequencies. The standard error is smaller than the symbol size. The inset illustrates the microscopic process leading to jets, in which two atoms collide, absorb one quantum from the oscillating field, and are ejected in opposite directions. 
	\label{fig:intro}}
\end{figure*}


The interplay between spontaneous and stimulated scattering events underpins many interesting physical phenomena. In general, spontaneous events dominate for low scattering rates. When the scattering rate exceeds a threshold, stimulated processes can ``run away'', leading to exponential amplification of outgoing particles. As a result, the character of the emission dramatically changes. A well-known example is the laser, in which a sufficient rate of stimulated emission results in a coherent wave of photons. Here, we show that runaway stimulation of collective atom-atom scattering in a driven Bose-Einstein condensate causes it to emit a burst of matter-wave jets.

To observe jets we typically perform experiments on thin, pancake-shaped condensates of $30\,000$ cesium atoms (Fig.~\ref{fig:intro}a, methods). The condensates homogeneously fill a circle of typical radius $R=8.5$~$\mu$m in the horizontal plane while being tightly, harmonically confined vertically to a root-mean-squared radius of $0.5$~$\mu$m. The trap depth in all directions is sufficient to confine the condensates, but weak enough to allow ejected atoms to propagate nearly undisturbed. 

After loading the condensates into this trap we modulate the magnetic field near a Feshbach resonance\cite{Chin2010} at frequency $f$ which causes the $s$-wave scattering length $a$ of the atoms to oscillate, see Fig.~\ref{fig:intro}b. Throughout this work we maintain a small, positive average scattering length $a_{\mathrm{dc}}=5~a_0$, in terms of the Bohr radius $a_0$. For typical experiments we hold the modulation amplitude at a constant value $a_{\mathrm{ac}}$ for a duration $\tau$ before imaging the atomic density distribution.

For the first few milliseconds of modulation little change is observed, until suddenly the jets emerge and propagate radially away from the condensate, see Fig.~\ref{fig:intro}c. The specific pattern of jets appears to be random in each repetition of the experiment. Even so, the jets have similar angular widths, and jets often appear to be accompanied by a partner propagating in the opposite direction. Side-view images of the condensates indicate that atoms are predominantly ejected in the horizontal plane, see Fig.~\ref{fig:intro}d. All of these behaviors are observed throughout a wide range of frequencies $f=1\sim10~$kHz.

To understand the microscopic process responsible for ejection of atoms, we extract the kinetic energy per atom by monitoring their distance from the condensate over time. We find that each atom has half of a quantum of the oscillating field, $E_\mathrm{k}=hf/2$ where $h$ is the Planck constant (Fig.~\ref{fig:intro}e). This relationship indicates that the ejected atoms come from collisions in which two atoms absorb and equally share an energy of $hf$ from the modulation and are ejected in opposite directions. From this microscopic perspective, the situation is similar to collisions between two condensates, during which counter-propagating pairs of atoms are ejected  while conserving momentum and energy\cite{Chikkatur2000}.

The preferential emission in the horizontal plane and the jet structure are salient and indicative of a collective collision process occurring throughout the condensate; uncorrelated, $s$-wave collisions should generate a diffuse, spherical shell of outgoing atoms. The observed features suggest that atoms produced in each collision stimulate further scattering into the same outgoing direction. Sufficient driving makes this stimulation run away and causes large numbers of atoms to go in particular directions and appear as jets while other directions have far fewer atoms and appear nearly empty. Note that the small, constant average scattering length is important for suppressing elastic collisions which scatter the atoms out of the jets. With sufficiently large positive or negative average scattering lengths $a_\mathrm{dc}$ we no longer observe jets. Furthermore, the disk shape of the condensates precludes strong stimulation for atoms emitted vertically, which rapidly escape the condensate before stimulating further collisions. This anisotropy results in the predominantly horizontal emission which we observe.

\begin{figure}
	\centering
	\includegraphics{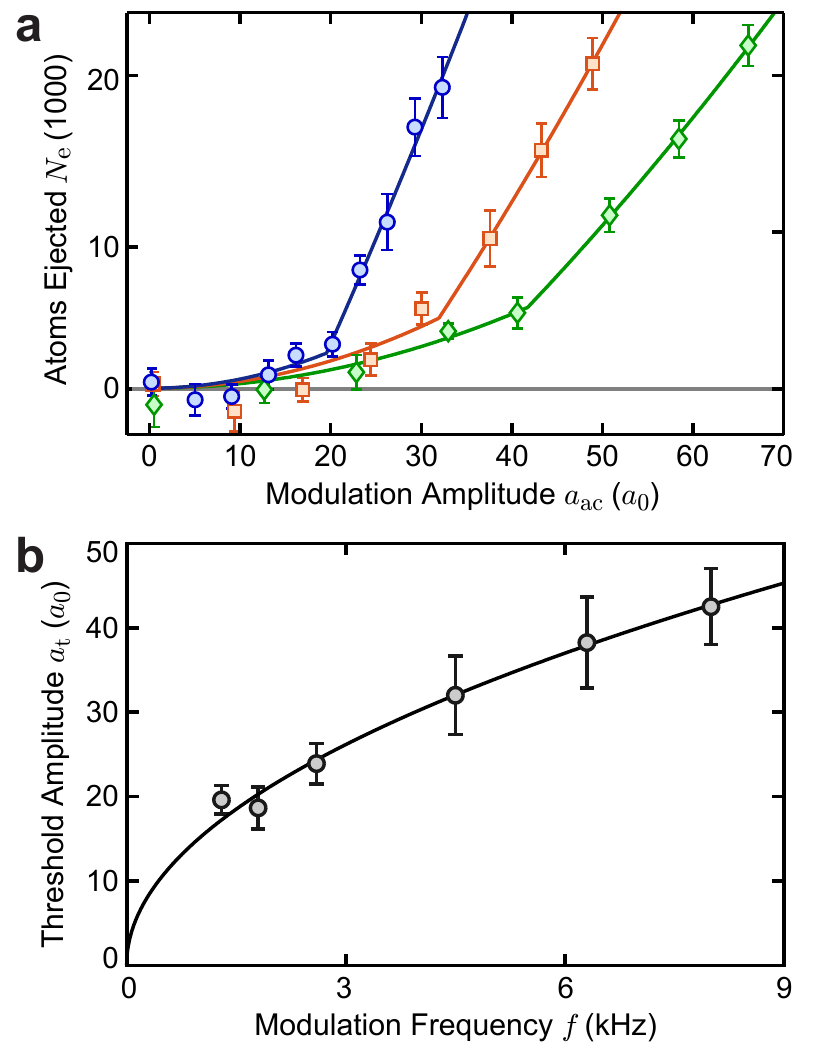} 
	\caption{\textbf{Threshold for jet formation.} 
		\textbf{a.} A plot of the number of ejected atoms $N_\mathrm{e}$ for modulation at frequency $f=1.3$~kHz for duration $\tau=46$~ms (blue circles), $f=4.5$~kHz for $\tau=25$~ms (orange squares), and $f=8.2$~kHz for $\tau=19$~ms (green diamonds). 
		Solid curves show fits based on Eq.~\ref{eqn:2a} (methods).  \textbf{b.} Threshold amplitudes extracted from fits as shown in panel \textbf{a}. The solid curve shows a fit based on Eq.~\ref{eqn:a_threshold} which yields the value of the numerical coefficient $\alpha=2.1(1)$. All error bars show standard errors. 
		\label{fig:threshold}}
\end{figure}

Runaway stimulated scattering only occurs when outgoing atoms stimulate further collisions faster than they escape the condensate. We estimate that the ejected atoms escape at a rate $\Gamma = \alpha \frac{v}{R}$ where 

\begin{equation}
v=\sqrt{\frac{hf}{m}}
\label{eqn:velocity}
\end{equation}

\noindent is the velocity of an ejected atom with mass $m$ and $\alpha$ is a dimensionless constant of order unity\cite{Vardi2002}. Moreover, a careful theoretical treatment yields an excitation rate for the ejected population $\gamma=\frac{2hna_\mathrm{ac}}{m}$ which is proportional to the modulation amplitude and the density $n$ of the condensate (see Supplementary material). Therefore, runaway stimulation occurs if the gain is larger than the loss $\gamma>\Gamma$.  When only the modulation amplitude is varied, we can recast the threshold condition as $a_\mathrm{ac}>a_\mathrm{t}$ where
\begin{equation}
a_\mathrm{t} = \alpha \frac{m}{2 h}\frac{v}{Rn}
\label{eqn:a_threshold}
\end{equation}

\noindent is the threshold amplitude. 

\begin{figure}
	\centering
	\includegraphics{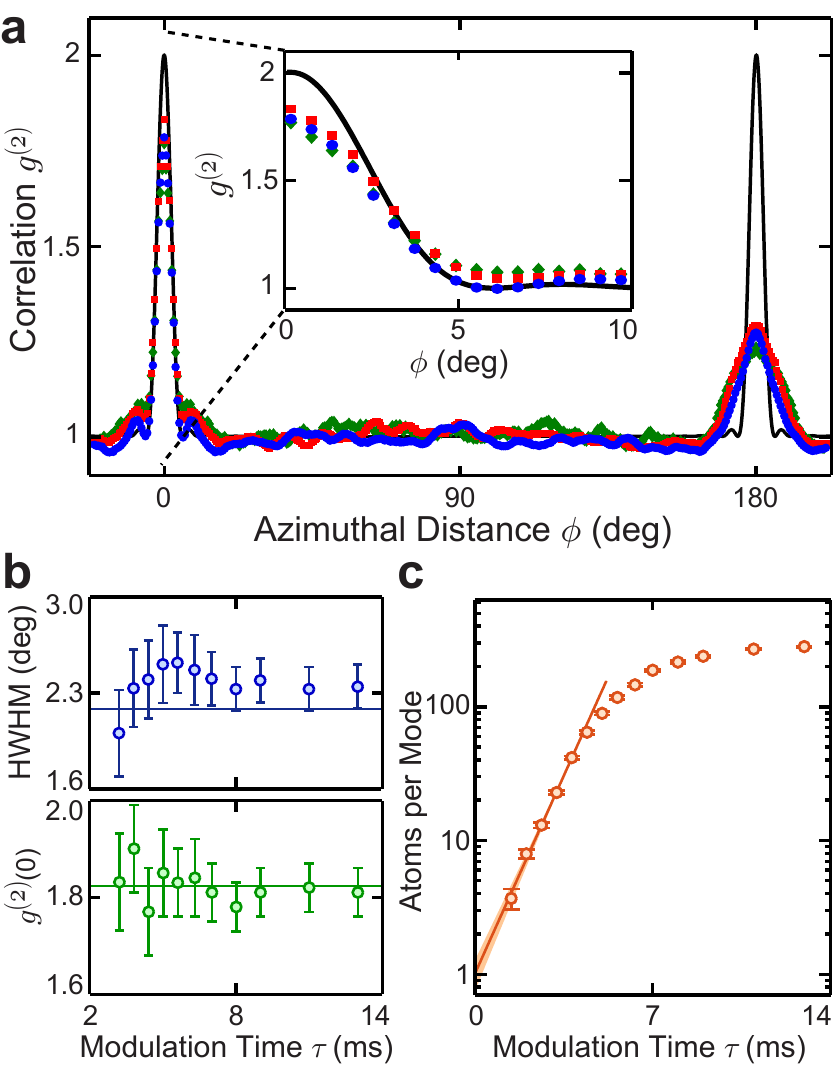} 
	\caption{\textbf{Correlations of emitted jets.} \textbf{a.} Azimuthal density-density correlations calculated from 90 images of jets emitted from condensates driven at $f=1.9$~kHz and $a_{\mathrm{ac}}=60~a_0$  for modulation durations of 4.4~ms (green diamonds), 5.6~ms (red squares), and 8.0~ms (blue circles). The solid curve shows the theoretical correlation function based on Eq.~\ref{eqn:g2theory}. The inset magnifies the peak near $\phi=0^\circ$. \textbf{b.} The half-width at half-maximum (HWHM, top) and height (bottom) of the peak near $\phi=0^\circ$ are constant within error bars, which represent one standard error. The solid lines represent the theoretical width (top) and the average measured height (bottom). \textbf{c.} The number of atoms per mode (circles) grows rapidly over time before saturating when the condensate is depleted. An exponential fit to the first five data points (solid curve) extrapolates to an initial level of 1.0(3) atoms per mode; the shaded region covers one standard error. 
		\label{fig:corrVsTime}}
\end{figure}

To test for the existence of a threshold for jet formation we measured the number of atoms ejected from the condensate at many different modulation amplitudes, see Fig.~\ref{fig:threshold}a. Below the threshold only a few atoms are ejected in a diffuse cloud, primarily from spontaneous scattering. Above a certain amplitude, which we identify as the threshold amplitude, the condensate suddenly starts ejecting many more atoms in the form of narrow jets. We further test the predicted behavior of the threshold (Eq.~\ref{eqn:a_threshold}) by varying the modulation frequency to change the velocity of outgoing atoms (Eq.~\ref{eqn:velocity}), see Fig.~\ref{fig:threshold}b. Our results verify the expected dependence $a_\mathrm{t}\propto\sqrt{f}$ and yield an empirical value $\alpha=2.1(1)$.

Many essential features of the jets are characterized by their correlations\cite{Pu2000, Duan2000, Vardi2002,Bach2002,Zin2005,Norrie2005,Ogren2009,Deuar2014,Wasak2014,Moore1999}. Specifically, we calculate the angular correlation function, 
\begin{equation}
g^{(2)}(\phi)=\frac{\braket{\int d\theta  n(\theta)\left[n(\theta+\phi)- \delta(\phi)\right]}}{\braket{\int d\theta n(\theta)}^2},
\label{eqn:g2definition}
\end{equation}
\noindent  where $n(\phi)$ is the angular density of atoms emitted at an angle $\phi$ and the angle brackets denote averaging over ensembles of many images (methods), see Fig.~\ref{fig:corrVsTime}a. We consistently detect two peaks in the measured correlation functions, one near $\phi=0^\circ$ and the other near $\phi=180^\circ$. The peak near $\phi=0^\circ$ results from collectively stimulated collisions, which lead to preferential bunching of the ejected atoms into the same modes. The second peak near $\phi=180^\circ$ appears because forward and backward jets are mutually stimulating as a result of conservation of momentum in the underlying pair scattering process. 

We derive the theoretical correlation function by calculating the time evolution of the excited, outgoing modes assuming they are initially in the vacuum state and using the Bogoliubov approximation for the condensate\cite{Pu2000, Duan2000, Vardi2002, Bach2002, Zin2005} (see Supplementary material). Our system is particularly amenable to this treatment because the condensate can be considered homogeneous and stationary.
From this treatment the correlation function is given by (see Supplementary material):
\begin{equation}
g_{\mathrm{th}}^{(2)}(\phi)= 1+\left|\frac{2J_{1}(k_f R \phi)}{k_f R \phi}\right|^2+\left|\frac{2J_{1}(k_f R (\phi-\pi))}{k_f R (\phi-\pi)}\right|^2,
\label{eqn:g2theory}
\end{equation}
\noindent where $J_n$ are the Bessel functions of the first kind and $k_f\equiv\sqrt{m\omega/\hbar}$ is the wavenumber of an outgoing atom. The profiles of the peaks reflect the angular profile of a jet which comes from the Fourier transform of the condensate density profile. The measured peaks near $\phi=0^\circ$ are in close agreement with the prediction for runaway stimulation, see Fig.~\ref{fig:corrVsTime}a inset. The slight reduction in the height of $g^{(2)}(0)$ is likely a result of our finite imaging resolution and a small amount of spontaneous emission into non-horizontal modes.

In addition to the jet profile, the correlation function also indicates the variance $\sigma^2=\left[g^{(2)}(0)-1\right]\braket{N_\theta}^2+\braket{N_\theta}$ (Eq.~\ref{eqn:g2definition}) of the atom number $N_\theta$ ejected in a direction $\theta$. For runaway stimulation and many ejected atoms, the standard deviation $\sigma\propto{N}$ is proportional to the number of ejected atoms, supporting the observed jet-like appearance. This contrasts with the case of spontaneous scattering, where the fluctuation $\sigma\propto\sqrt{N}$ comes only from shot noise, leading to a diffuse halo. 

Conservation of momentum in pair scattering should in principle cause each jet to be accompanied by a counter-propagating partner, such that the peak in $g^{(2)}$ near $\phi=180^\circ$ has the same area as the peak near $\phi=0^\circ$. Taking the ratio of the areas we obtain $A(180^\circ)/A(0^\circ)=70\sim85\%$, suggesting that this expectation is largely met. However, the peaks near $\phi=180^\circ$ are shorter and wider than those near $\phi=0^\circ$. We note that the profiles of the peaks near $\phi=180^\circ$ are much more sensitive to technical distortions of the atom trajectories, since this peak comes from atoms on opposite sides of the image which are separated by approximately 150~$\mu$m at the time of detection. In addition, broadening of the correlation peaks has been predicted for analogous system\cite{Ogren2009}. Further investigation into the differences between the two peaks is required. 

\begin{figure}
	\centering
	\includegraphics{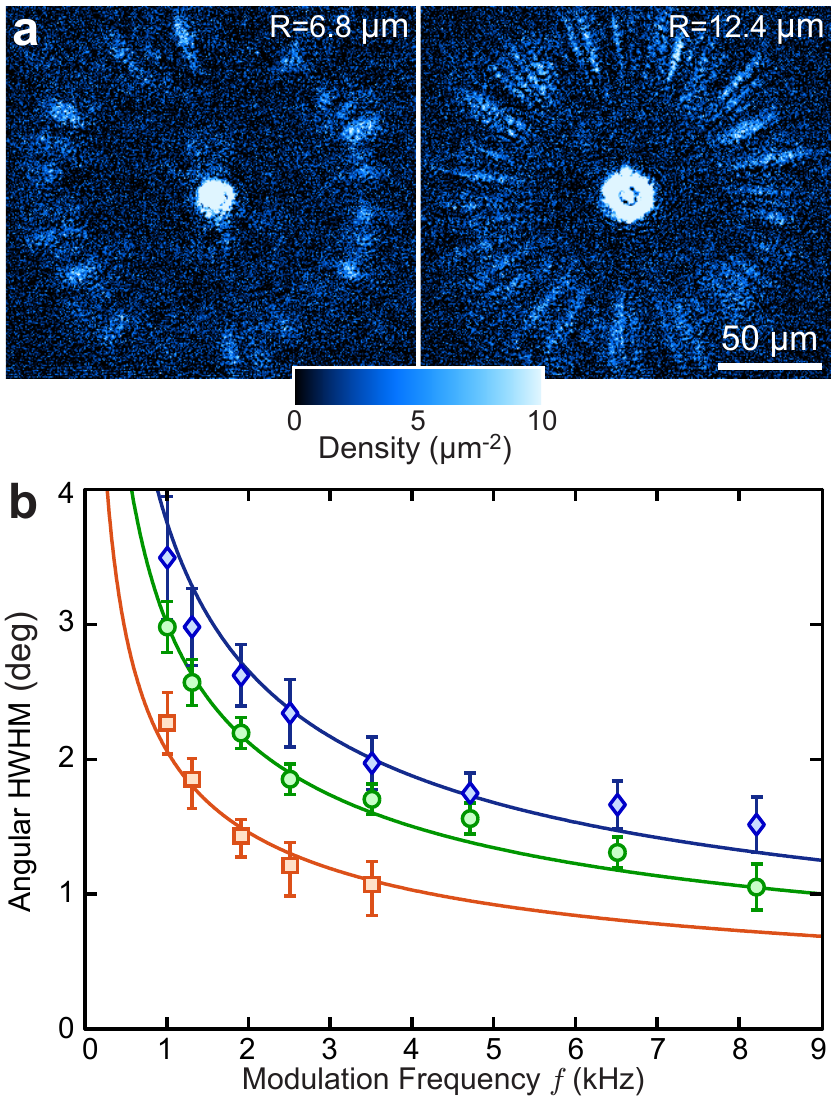} 
	\caption{\textbf{Angular width of the jets.} 
	\textbf{a.} Example images of jets ejected from condensates of different radii with modulation frequency $f=1.9$~kHz.
 	\textbf{b.} The deconvolved widths of the peaks at $g^{(2)}(0)$ for condensates with radius $R=6.8$ (blue diamonds), $8.5$ (green circles), and $12.4$~$\mu{m}$ (orange squares). Solid curves indicate the theoretical widths based on Eq.~\ref{eqn:g2theory}. Error bars indicate one standard error. For details on experimental parameters refer to the methods.
		\label{fig:corrVsRf}}
\end{figure}

The correlation function is remarkably consistent throughout the amplification process even as the number of atoms in jets grows by an order of magnitude, see Fig.~\ref{fig:corrVsTime}. When the modulation duration varies from $\tau=2\sim14$~ms we find that both the width and height of the zero peak remain constant. This observation is consistent with the expectation that the gain in our system is isotropic in the horizontal plane; as a result, the stimulated emission into each horizontal mode depends only on its occupancy and the runaway stimulation does not change the jet structure.

Since the correlations appear to be time-independent, the distribution of atoms in the amplified jets should reflect the fluctuation of the original condensate. The thermal fluctuations are expected to be negligible since the thermal population $\frac{1}{e^{(E-\mu)/kT}-1}\leq0.04$ atoms per outgoing mode is small for our condensate with chemical potential of $\mu= h\times25~$Hz and temperature of $T=7$~nK modulated at frequencies $f\geq1~$kHz. As a result, quantum fluctuation should dominate the initial population, suggesting that we observe parametric amplification of the vacuum fluctuation. To further probe this feature we calculate the apparent number of atoms per outgoing mode as a function of the driving time, see Fig.~\ref{fig:corrVsTime}c. Here we estimate the number of outgoing modes as $M\approx180^\circ/\Delta{\theta}=78$ where $\Delta{\theta}=2.3^\circ$ is the measured half-width at half-maximum of the peak at $g^{(2)}(0)$. The growth of the atom number over time is well described by an exponential until the depletion of the condensate becomes relevant. An exponential fit extrapolates to an apparent initial number of $1.0(3)$; in the case of quantum fluctuation this value represents a virtual population.

An important prediction is that the width of the jets has a simple relationship to the size of the condensate and the modulation frequency by the Heisenberg uncertainty principle, see Fig.~\ref{fig:corrVsRf}. In particular, the uncertainty principle suggests that the lower limit of the transverse spread of momentum in a jet $\Delta{k}\propto 1/R$ is set by the radius of the condensate, and the radial momentum $k_f\propto1/\sqrt{f}$ is determined by the oscillation frequency. As a result, the uncertainty-limited angular width follows $\Delta{\theta}=\Delta{k}/k_f\propto1/(Rk_f)$. The complete calculation, see Eq.~\ref{eqn:g2theory}, yields $\Delta{\theta}\approx1.62/(Rk_f)$. The measured angular widths of jets emitted from condensates of various sizes and oscillation frequencies saturate this uncertainty limit, see Fig.~\ref{fig:corrVsRf}.
Interestingly, the angular width of the jets does not appear to depend on their distance from the condensate. 

The jet emission process can be further investigated as a function of the coherence length, shape, or collective motion of the gas. Alternatively, jets could serve as a probe of the excitations present in Bose condensates. Moreover, this system may provide an intriguing approach to generating twin matter waves for metrological applications\cite{Bookjans2011,Lucke2011,Gross2011}. Jet emission should also be taken into account in schemes for Floquet engineering which utilize oscillating interactions\cite{Rapp2012,Meinert2016}.

We thank Erez Berg for helpful discussions. L.~W.~C. is supported by the Grainger graduate fellowship. A.~G. is supported by the Kadanoff-Rice fellowship. This work is supported by NSF Materials Research Science and Engineering Centers (DMR-1420709), NSF grand PHY-1511696, and Army Research Office-Multidisciplinary Research Initiative grant W911NF-14-1-0003. The data presented in this paper are available upon request to C.~C. (cchin@uchicago.edu). \\

\noindent\textbf{Methods}

\noindent\textbf{Condensate Preparation} 
Horizontal confinement is provided by a repulsive $780$~nm laser which is shaped into a circle of controllable radius by a digital micromirror device (DMD) before being projected onto the atoms through a high-resolution objective lens (numerical aperture NA=0.5). The resulting barrier has a height of $h\times150$~Hz and thickness of $4$~$\mu$m. Vertical ($z$-axis) confinement is approximately harmonic with frequency $\omega_z=2\pi\times210$~Hz and depth $h\times500$~Hz. 

Note that the condensates of $R=6.8$~$\mu$m and $12.4$~$\mu$m used in Fig.~\ref{fig:corrVsRf} typically contain $N=26\,000$ and $41\,000$ atoms, respectively. To control the scattering length we modulate the magnetic field around $17.22$~G, near the zero-crossing of a Feshbach resonance\cite{Chin2010}. 

\noindent\textbf{Threshold Measurement} 
To determine the threshold at each frequency, we measure the number of atoms ejected as a function of the modulation amplitude. Here, the duration $\tau$ is the time an atom takes to travel $80~\mu$m to the edge of our field of view. We fit the results using an empirical model,
	\begin{equation}
	N_\mathrm{e} = A a_\mathrm{ac}^2 + B (a_\mathrm{ac}-a_\mathrm{t})\Theta(a_\mathrm{ac}-a_\mathrm{t}),
	\label{eqn:2a}
	\end{equation}
	where $A$ and $B$ characterize the strength of the spontaneous and strongly stimulated contributions to emission, respectively, and $\Theta$ is the Heaviside step function.

\noindent\textbf{Correlation Functions} 
We calculate the angular correlation functions (Eq.~\ref{eqn:g2definition}) from our images using discrete angular slices of width 10~mrad. For each condition we include only atoms within an annulus whose inner and outer radii are symmetric around the distance at which ejected atoms are most dense. For Fig.~\ref{fig:corrVsTime} the annulus has thickness $10$~$\mu$m, whereas for Fig.~\ref{fig:corrVsRf} the thickness increases to $20$~$\mu$m to improve the signal strength for conditions with few jets. The half-widths at half-maximum shown in Figs.~3~and~4 are corrected for small systematic shifts due to our finite imaging resolution of $1.4$~$\mu$m\cite{Hung2011}.

\section{Supplementary Information: Theory}

The many-body Hamiltonian describing our system is,
	
	\begin{align*}
	H=&\int d^{3}\boldsymbol{r}\Psi^{\dagger}(\boldsymbol{r,}t)\frac{p^{2}}{2m}\Psi(\boldsymbol{r},t)+\int d^{3}\boldsymbol{r}\Psi^{\dagger}(\boldsymbol{r,}t)V(\boldsymbol{r})\Psi(\boldsymbol{r},t)\\
	&+\frac{g(t)}{2}\int d^{3}\boldsymbol{r}\Psi^{\dagger}(\boldsymbol{r,}t)\Psi^{\dagger}(\boldsymbol{r,}t)\Psi(\boldsymbol{r},t)\Psi(\boldsymbol{r},t)
	\end{align*}
	
	\noindent where $\Psi$ is the bosonic field operator, $V(\boldsymbol{r})$
	is a static external potential which determines the initial shape
	of the condensate, and $m$ is the mass. Dynamics are driven by an oscillating interaction
	strength $g(t)=\frac{4\pi\hbar^{2}a(t)}{m}$ where we recall that the scattering length follows
	$a(t) = a_{\mathrm{dc}}+a_{\mathrm{ac}}\mathrm{sin}(\omega t)$ as in the main text. 
	
	We begin by invoking a few key assumptions. First, since most of this work
	can be understood in the regime where the depletion of the condensate
	is negligible, we will use the Bogoliubov approximation of a fixed,
	macroscopically occupied condensate. Therefore, we decompose the field operator
	into a quantum field describing the excited modes $\Psi_{e}(\boldsymbol{r})$
	and a classical field $\psi_{0}(\boldsymbol{r},t)$ which corresponds to the condensate wavefunction. Second,
	since we do not observe significant ejection of atoms outside the
	horizontal plane, we will assume that the emission is only into horizontal
	modes and ignore the vertical structure of the gas. Further assuming an idealized
	trap and ignoring the healing length, we can approximate the condensate
	as a homogeneous cylinder of radius $R$ and density $n$, such that
	the classical field describing the condensate is $\psi_{0}(\boldsymbol{r})\equiv\sqrt{n}\rho(\boldsymbol{r})$
	where
	\[
	\rho(\boldsymbol{r})=\begin{cases}
	1 & r\leq R\\
	0 & r>R
	\end{cases}
	\]
	in cylindrical coordinates with the horizontal radius $r$. Since
	the trap potential $V(r)$ primarily serves to contain the condensate
	and does not significantly affect jet propagation, we will neglect
	its effects on the time evolution of the excited modes. Moreover,
	it is convenient to work in Fourier space, using the transformations %

	\[
	\rho(\boldsymbol{r})=\frac{1}{(2\pi)^{3/2}}\int d^{3}ke^{i\boldsymbol{k\cdot r}}\tilde{\rho}(\boldsymbol{k}),
	\]
	\[
	\Psi_{e}(\boldsymbol{r},t)=\frac{1}{(2\pi)^{3/2}}\int d^{3}ke^{i\boldsymbol{k\cdot r}}b(\boldsymbol{k},t).
	\]

	We can now invoke the rotating wave
	approximation. We define $b(\boldsymbol{k},t)\equiv e^{-i\omega t/2}c(\boldsymbol{k},t)$.
	Time evolution of the rotating operators $c(\boldsymbol{k},t)$ is
	governed by the Hamiltonian $H_{c}=H-\int d^{3}k\frac{\hbar\omega}{2}c^{\dagger}(\boldsymbol{k},t)c(\boldsymbol{k},t)$ in which we drop terms which oscillate at multiples of $\omega$. We also drop the elastic scattering term proportional to $a_\mathrm{dc}$, which is negligible in this work.
	Together, these steps yield the Hamiltonian: 
	
	\begin{align*}
	H_{c} & =\frac{\hbar^{2}}{2m}\int d^{3}\boldsymbol{k}(k^{2}-k_{f}^{2})c^{\dagger}(\boldsymbol{k})c(\boldsymbol{k})\\
	& +\frac{\hbar\gamma}{4}\int d^{3}k_{1}d^{3}k_{2}\left(c^{\dagger}(\boldsymbol{k}_{1})c^{\dagger}(\boldsymbol{k}_{2})\frac{\tilde{\rho}(\boldsymbol{k_{1}+k_{2}})}{(2\pi)^{3/2}}+\mathrm{h.c.}\right)
	\end{align*}
	where the excitation rate is
	
	\[
	\gamma=\frac{2hna_{\mathrm{ac}}}{m}.
	\]
	
	\noindent Below, in the absence of the decay term, the excitation rate will appear as the rate of exponential growth for the excited atom number. 
	
	The time evolution of the excited field in the Heisenberg representation yields,

	\[
	\dot{c}(\boldsymbol{k})=-i\frac{\hbar}{2m}(k^{2}-k_{f}^{2})c(\boldsymbol{k})-i\frac{\gamma}{2}\int d^{3}k_{1}\frac{\tilde{\rho}(\boldsymbol{k+k_{1}})}{(2\pi)^{3/2}}c^{\dagger}(\boldsymbol{k_{1}}),
	\]
	\noindent where $k_{f}\equiv\sqrt{m\omega m/\hbar}$ is the carrier wavenumber. The first term contains 			the kinetic energy of the excited atoms and
	the second term contains the interactions which populate the excited
	modes. We expect the dynamics to be dominated by wavepackets whose
	carrier wavenumber $k=k_f$ is on resonance and whose envelope takes the shape
	of the condensate. In this case, the first term encodes the motion
	of the wavepacket, which leads to the decay of population from the
	condensate and therefore causes the threshold behavior. 
	
	For the purpose of
	calculating the correlation function, let us assume that the system
	is sufficiently above the threshold that the kinetic energy term is negligible and
	focus on the interaction term. Furthermore, notice that the integral effectively performs a projection
	onto the homogeneous condensate density profile $\rho(\boldsymbol{r})$; that is, $\int d^{3}k_{1}\frac{\tilde{\rho}(\boldsymbol{k+k_{1}})}{(2\pi)^{3/2}}c^{\dagger}(\boldsymbol{k_{1}})=c_{in}^{\dagger}(-\boldsymbol{k})$
	where the subscript ``in'' denotes a projection onto the inside
	of the condensate boundary and we can decompose the operator as $c(\boldsymbol{k})=c_{\mathrm{out}}(\boldsymbol{k})+c_{\mathrm{in}}(\boldsymbol{k})$.
	The evolution equation is then 
	\[
	\dot{c}_{\mathrm{in}}(\boldsymbol{k})+\dot{c}_{out}(\boldsymbol{k})=-i\frac{\gamma}{2}c_{in}^{\dagger}(\boldsymbol{-k}).
	\]
	This equation has the solution
	\[
	c_{out}(\boldsymbol{k},t)=c_{out}(\boldsymbol{k},0),
	\]
	
	\begin{equation}
	c_{in}(\boldsymbol{k},t)=c_{in}(\boldsymbol{k},0)\mathrm{cosh}(\mathit{\frac{\gamma}{2}t})\mathrm{-\mathit{i}\mathit{c_{in}^{\dagger}}(-\mathit{\boldsymbol{k}},0)\mathrm{sinh(\mathit{\frac{\gamma}{2}t})}},
	\label{eqn:time_evolution}
	\end{equation}
	
	\noindent in which all amplitudes inside the condensate boundary grow, and those
	outside do not. The thermal fluctuation at the relevant energies is
	expected to be negligible, so we can assume that the initial state
	is the vacuum and all of the excited modes are initially empty. Therefore
	we have 
	\[
	c(\boldsymbol{k}_{1},0)\left|0\right\rangle =0
	\]
	\[
	\left\langle c(\boldsymbol{k}_{1},0)c^{\dagger}(\boldsymbol{k_{2}},0)\right\rangle =\delta(\boldsymbol{k_{1}-k_{2}}).
	\]
	
	The correlation function is defined as
	\[
	g^{(2)}(\boldsymbol{k_{1},}\boldsymbol{k_{2}},t)\equiv\frac{\left\langle c^{\dagger}(\boldsymbol{k_{1}},t)c^{\dagger}(\boldsymbol{k_{2}},t)c(\boldsymbol{k_{2}}t)c(\boldsymbol{k_{1}},t)\right\rangle }{\left\langle n(\boldsymbol{k_{2}},t)\right\rangle \left\langle n(\boldsymbol{k_{1}},t)\right\rangle }.
	\]
	Using Wick's theorem, which applies because the time evolution is linear (Eq.~\ref{eqn:time_evolution}), we can rewrite the correlation function as,
	
	\[
	g^{(2)}(\boldsymbol{k_{1},k_{2}},t)=1+\frac{\left|n(\boldsymbol{k_{1},k_{2},}t)\right|^{2}+\left|m(\boldsymbol{k_{1},k_{2},}t)\right|^{2}}{\left\langle n(\boldsymbol{k_{2}},t)\right\rangle \left\langle n(\boldsymbol{k_{1}},t)\right\rangle },
	\]
	
	where $n(\boldsymbol{k_{1},k_{2},}t)=\left\langle c^{\dagger}(\boldsymbol{k_{1}},t)c(\boldsymbol{k_{2}},t)\right\rangle $
	is the contribution from the density matrix, we have defined $n(\boldsymbol{k},t)\equiv n(\boldsymbol{k,k},t)$,
	and $m(\boldsymbol{k_{1},k_{2},}t)=\left\langle c(\boldsymbol{k_{1},}t)c(\boldsymbol{k_{2}},t)\right\rangle $
	is the anomalous contribution. Substituing the solution for the excited field from Eq.~(\ref{eqn:time_evolution}), we obtain,

	\[
	n(\boldsymbol{k_{1},k_{2},}t)=\frac{\tilde{\rho}(\boldsymbol{k_{1}-k_{2}})}{(2\pi)^{3/2}}\mathrm{sinh}^{2}(\mathit{\frac{\gamma}{2}t}),
	\]
	\\
	for the density matrix. Note that $\mathrm{sinh}^{2}(\mathit{\frac{\gamma}{2}t})$
	contains one term which is exponentially growing at a rate $\gamma$. Accounting for the effects of the kinetic energy term, which leads to the threshold behavior discussed in the main text, yields a reduced exponential growth rate satisfying $\gamma'=2hn(a_\mathrm{ac}-a_\mathrm{t})/m$, consistent with our observations in Fig.~3c of the main text. 
	
	Furthermore, far above the threshold we obtain
	
	\[
	m(\boldsymbol{k_{1},k_{2},}t)=-i\frac{\tilde{\rho}(\boldsymbol{k_{1}+k_{2}})}{(2\pi)^{3/2}}\mathrm{cosh}(\mathit{\frac{\gamma}{2}t})\mathrm{sinh(\mathit{\frac{\gamma}{2}t})}
	\]
	for the anomalous contribution. Combining these results, we find the correlation function
	\begin{align*}
	g^{(2)}(\boldsymbol{k_{1},k_{2}},t)=1&+\frac{\left|\tilde{\rho}(\boldsymbol{k_{1}-k_{2}})\right|^{2}}{\tilde{\rho}(0)^{2}}\\
	&+\frac{\left|\tilde{\rho}(\boldsymbol{k_{1}+k_{2}})\right|^{2}}{\tilde{\rho}(0)^{2}}\mathrm{coth}^{2}(\frac{\gamma}{2}t).
	\end{align*}

	We can simplify the correlation function by assuming that $\left|\boldsymbol{k}_{1}\right|=\left|\boldsymbol{k}_{2}\right|=k_{f}$
	and calculating the correlations as a function of the angle $\phi$
	between $\boldsymbol{k}_{1}$ and $\boldsymbol{k}_{2}$. Furthermore,
	taking into account the disk-shape of our condensates and noting that
	our observations are made at times satisfying $\gamma t\gg1$,
	we obtain the time-independent correlation function:
	
	\[
	g^{(2)}(\phi)=1+\left|\frac{2J_{1}(k_{f}R\phi)}{k_{f}R\phi}\right|^{2}+\left|\frac{2J_{1}(k_{f}R[\phi-\pi])}{k_{f}R(\phi-\pi)}\right|^{2},
	\]
	where in our experiments the coefficient $k_{f}R\gg1$ is sufficiently large 
	that we have made small angle approximations for the last two terms.

\end{document}